\documentclass[a4paper,onecolumn,11pt]{quantumarticle}
\pdfoutput=1
\usepackage[utf8]{inputenc}
\usepackage[english]{babel}
\usepackage[T1]{fontenc}
\usepackage{amsmath}
\usepackage{hyperref}
\usepackage{braket}
\usepackage{tikz}
\usepackage{lipsum}
\usepackage[numbers]{natbib}

\usepackage{graphicx}
\usepackage{dcolumn}
\usepackage{bm}
\usepackage{color} 

\newcommand{\beq}{\begin{equation}}
\newcommand{\eeq}{\end{equation}}
\newcommand{\beqa}{\begin{eqnarray}}
\newcommand{\eeqa}{\end{eqnarray}}

\begin{document}
\title{\parbox{\linewidth}{Impact of Force Noise on the Visibility of STA-Based Atom Interferometers}}
\author{S. Mart\'\i nez-Garaot}
\affiliation{Department of Physical Chemistry, University of the Basque Country UPV/EHU, Apartado 644, Bilbao 48080, Spain}
\affiliation{EHU Quantum Center, University of the Basque Country UPV/EHU, 48940 Leioa, Spain}
\orcid{0000-0002-6916-3858}
\author{I. Lizuain}
\affiliation{Department of Applied Mathematics, University of the Basque Country UPV/EHU, Bilbao 48013, Spain}
\affiliation{EHU Quantum Center, University of the Basque Country UPV/EHU, 48940 Leioa, Spain}
\orcid{0000-0001-9207-4493}
\author{A. Rodriguez-Prieto}
\affiliation{Department of Applied Mathematics, University of the Basque Country UPV/EHU, Bilbao 48013, Spain}
\affiliation{EHU Quantum Center, University of the Basque Country UPV/EHU, 48940 Leioa, Spain}
\orcid{0000-0002-9030-0060}
\author{J. G. Muga}
\affiliation{Department of Physical Chemistry, University of the Basque Country UPV/EHU, Apartado 644, Bilbao 48080, Spain}
\affiliation{EHU Quantum Center, University of the Basque Country UPV/EHU, 48940 Leioa, Spain}
\orcid{0000-0002-1967-502X}
%
%
%

\maketitle
\begin{abstract}
%
In a previous work \cite{Rodriguez-Prieto2020}, we designed a compact atom interferometer to measure homogeneous constant forces guiding the arms via shortcuts to adiabatic paths. Within this scheme we drive the atom by moving spin-dependent traps, and design a force $f(t)$  to compensate inertial terms in the moving frame. In this paper we analyze  how robust our interferometer is against some realistic noisy deviation from the unperturbed value of the force $f(t)$. The complex overlap of the atom wave functions for each arm of the interferometer at the final time of the process is calculated. We demonstrate that the measure of the unknown force  will not be affected, as there will be no contribution of the error in the phase difference of the overlap. Nevertheless, the visibility will be reduced as the modulus will no longer be one. In addition we find the optimal trajectories minimizing the effect of the noise in the visibility while keeping the sensitivity of the interferometer.
\end{abstract}

\section{Introduction}
Atom interferometry \cite{berman1997, Cronin2009}  provides a route to quantum enhanced precise sensors, including inertial detectors with impressive sensitivities, and high-precision measurements in gravimeters, gyrometers and velocity sensors. The key idea is to split and later recombine the atomic wavefunction, whose interference pattern is sensitive to the differential phase accumulated during the separation. 

In a previous paper \cite{Rodriguez-Prieto2020} we worked out a scheme to measure constant homogeneous forces using STA techniques to achieve adiabatic arm guiding \cite{Martinez-Garaot2018, Dupont-Nivet2016, Navez2016, Palmero2017}. Here, STA stands for ``Shortcuts To Adiabaticity'', a set of techniques designed to obtain the results of adiabatic dynamics in significantly shorter times \cite{Torrontegui2013, Guery-Odelin2019}, whereas by ``guide'' we mean that the atom is driven with two oppositely moving traps and follow a previously designed trajectory. Specifically, we guide the atom via optical lattices to accelerate the arm wavefunctions for a single internal state \cite{Muller2009,Mandel2003}, which allows precise measurements at the ultrashort spatial scale \cite{Kovachy2010, Steffen2012}. Furthermore, our STA-mediated guided interferometer scheme realizes the ideal aim of giving a motional-state independent differential phase, while simultaneously keeping a high sensitivity, and within short process times. 

Any scheme proposed for high-precision measurements must fulfill certain requirements, such as robustness against different types of noises. 
In this work, besides revisiting our proposed scheme, we will consider a realistic noisy deviation from the unperturbed value of the $f(t)$ force. Within our interferometer this force plays a fundamental role as it is inverse engineered in order to compensate the trap acceleration. Such a compensation is one of the best ways to implement STA-driven fast transport \cite{Torrontegui2011}. 
To model the effect of noise in our system, we adopt a perturbative approach similar to that developed by Lu et al. \cite{Lu2014, Lu2018, Lu2020,Ruschhaupt2012}, where the impact of weak classical noise on STA-based transport is analyzed in terms of noise sensitivities. In our case, we consider fluctuations in the compensating force f(t), which enter the Hamiltonian as a time-dependent perturbation. Following this framework, we compute the overlap between the two interferometric branches up to second order in the noise strength. Importantly, we find that the phase difference—responsible for encoding the measured force—remains unaffected by the noise, while the modulus of the overlap is reduced. This leads to a degradation of the visibility, which is directly linked to the loss of coherence between the arms. We quantify this effect analytically and numerically. We, also, propose optimized trajectories that minimize the visibility loss due to the effect of the noise, while preserving the sensitivity of the interferometer. In order to do so we need to numerically solve a mathematical problem consisting in minimizing an objective functional with an integral isoperimetric constraint.

Recent developments have highlighted the importance of robust control techniques to mitigate the effects of noise in atom interferometry. For instance, tailored light pulses designed via robust quantum control have been shown to preserve interferometric sensitivity and scale factor even under strong laser-intensity fluctuations and platform-induced motion \cite{Saywell2023}. In parallel, enhanced shortcuts to adiabaticity (eSTA) have been proposed as a refinement of STA methods, enabling high-fidelity atom transport in complex, anharmonic, and multidimensional potentials such as double-well optical lattices \cite{Hauck2022}. These advances reinforce the relevance of designing control protocols that are both fast and resilient to noise, a central theme in our present analysis.

\section{The interferometer and traps movement}
\label{interferometer}
Let us first review the basic equations of our interferometer introduced in \cite{Rodriguez-Prieto2020}. 
For a single atom with two internal states (spin up, $\ket{\uparrow}$, and spin down, $\ket{\downarrow}$) and effective motion in one dimension, the atom state at time $t$ can be written as
\beq
\Psi(x,t)=a_\uparrow \ket{\uparrow} \psi^\uparrow(x,t)+a_\downarrow \ket{\downarrow} \psi^\downarrow(x,t), 
\eeq
with $\psi^{\uparrow \downarrow}(x,t)=\bra{x}\ket{\psi^{\uparrow \downarrow}(t)}$ the motional states for the two internal levels in coordinate representation. 
We assume a prepared state $\ket{\uparrow} \ket{\Phi_p}$ from which a $\pi/2$ pulse produces two equally weighted components with $a_\uparrow=a_\downarrow=\frac{1}{\sqrt{2}}$. Setting the initial time $t=0$ at the end of the $\pi/2$ pulse and $\Phi(x,0)\equiv \Phi_p(x)=\psi^\uparrow(x,0)=\psi^\downarrow(x,0)$, the population in each internal state at final time $t_f$ is
\beq
P^{\uparrow \downarrow}(t_f)=\frac{1}{2}\pm\frac{1}{2}Re[\bra{\psi^\downarrow(t_f)}\ket{\psi^{\uparrow}(t_f)}], 
\eeq
and the complex overlap can be written as
\beq
\label{overlap}
\bra{\psi^\downarrow(t_f)}\ket{\psi^\uparrow(t_f)}=e^{i \Delta\varphi(t_f)}|\bra{\psi^\downarrow(t_f)}\ket{\psi^\uparrow(t_f)}|.
\eeq
Notice that here, and in the following, the superscript $\uparrow \downarrow$ in any equation implies that the sign on top in $\mp$ or $\pm$ is for $\uparrow$, whereas the sign on the bottom is for $\downarrow$. 
To guarantees the modulus of (\ref{overlap}) to be $1$ or, in other words, to keep maximal visibility we guide the arms wavefunctions of the interferometer via shortcuts to adiabatic paths \cite{Rodriguez-Prieto2020}. In this scenario, the populations read $P^{\uparrow\downarrow}(t_f)=\frac{1}{2}\pm\frac{1}{2}\cos{[\Delta \varphi(t_f)]}$ and the phase difference is proportional to the unknown force, $\Delta \varphi(t_f)=S c$. Therefore, since the sensitivity $S$ is known, it is possible measure the unknown force $c$ through the measurement of the populations \cite{Rodriguez-Prieto2020, Martinez-Garaot2018}. 


Once the interferometer is presented, we will also make a brief summary of the method used to design the traps movement. Assuming a Lamb-Dicke regime and different evolution for each spin state, the Hamiltonian can be written as
\beq
\label{Hinitial}
H^{\uparrow \downarrow}=\frac{p^2}{2 m}-cx\mp [x-x_0(t)]f(t)+U[x\mp\alpha(t)],  
\eeq
where $x_0(t)$ is the ``crossing point'' of the potential energies for the spin-dependent forces, $c$ is the spin-dependent homogeneous-in-space and constant-in-time force we want to measure, $m$ is the mass of the atom and $p^2/2m$ is the Kinetic energy. 
The force $f(t)$ will be chosen to compensate inertial terms in the moving frame and the trap potentials $U[x\mp\alpha(t)]$ move along opposite trajectories $\alpha^{\uparrow \downarrow}(t)=\pm \alpha(t)$. We consider trap trajectories that satisfy the boundary conditions
\beq
\label{bc}
\alpha(t_b)=\dot\alpha(t_b)=0, 
\eeq
at the boundary times $t_b=0,t_f$. Here and throughout the paper dots denote time derivatives.
Defining 
%
\beqa
\tilde U[x\mp \alpha(t)]&=&U[x\mp\alpha(t)]-[x\mp\alpha(t)]c, \\
%
%
\Lambda^{\uparrow\downarrow}(t)&=&\pm f(t)x_0(t)\mp c \alpha(t),  
\eeqa
we can rewrite the Hamiltonian in (\ref{Hinitial}) as follows:
\beq
H^{\uparrow \downarrow}=\frac{p^2}{2m}\mp f(t)x+\tilde U (x\mp\alpha)+\Lambda^{\uparrow \downarrow}(t). 
\eeq

To solve the dynamics, it is useful to perform unitary transformations into ``moving-frame interaction pictures'' in which the interaction picture wave-vectors $\ket{\psi_I^{\uparrow \downarrow}}$ are defined in terms of Schr\"odinger (laboratory frame) wave-vectors $\ket{\psi^{\uparrow \downarrow}}$ as
\beqa
\label{IP}
\ket{\psi_I^{\uparrow \downarrow}}&=&\mathcal{U}^{\uparrow \downarrow} \ket{\psi^{\uparrow \downarrow}}, \nonumber \\
\ket{\psi^{\uparrow \downarrow}}&=&(\mathcal{U}^{\uparrow \downarrow})^\dagger \ket{\psi_I^{\uparrow \downarrow}},
\eeqa 
with the unitary operator 
\beq
\label{U}
\mathcal{U}^{\uparrow \downarrow}=e^{\pm i \alpha p/\hbar} e^{\mp im \dot\alpha x/\hbar}. 
\eeq
Taking into account Eqs. (\ref{IP}) and (\ref{U}) the effective moving-frame Hamiltonian becomes
\beqa
\label{HIP}
H^{\uparrow \downarrow}_I&=&\mathcal{U}^{\uparrow \downarrow} H^{\uparrow \downarrow} (\mathcal{U}^{\uparrow \downarrow})^\dagger + i \hbar \dot{\mathcal{U}}^{\uparrow \downarrow} (\mathcal{U}^{\uparrow \downarrow})^\dagger, \nonumber \\
&=&\frac{p^2}{2m}+\frac{1}{2}m \dot\alpha^2 \mp (x\pm \alpha) f(t)+ \tilde U(x) \pm f(t)x_0(t) \mp c \alpha \pm (x \pm \alpha) m \ddot \alpha. 
\eeqa
If the auxiliary force satisfies the Newton equation
\beq
\label{Newton}
\ddot \alpha(t)=\frac{f(t)}{m} 
\eeq
for the trajectory $\alpha(t)$ compensates for inertial effects due to the motion of the $\tilde U[x\mp \alpha(t)]$ potentials in the laboratory frame. The consequence is that a stationary state in the moving frame will remain so. 

In our design the force $f(t)$ is designed inversely from the trajectory $\alpha(t)$ according to Eq. (\ref{Newton}). Therefore, to make $f(t_b)=0$ we may to impose the additional boundary conditions $\ddot \alpha(t_b)=0$. Substituting Eq. (\ref{Newton}) in Eq. (\ref{HIP}), the moving-frame Hamiltonians take a simple form with a common time- and spin-independent term $H_{I,0}$ and terms $F^{\uparrow \downarrow} (t)$ that depend on time but not on $x$ or $p$, 
\beqa
H^{\uparrow \downarrow}_I&=&H_{I,0}+F^{\uparrow \downarrow} (t), \nonumber \\
H_{I,0}&=&\frac{p^2}{2m}+\tilde U(x), \nonumber \\ 
F^{\uparrow \downarrow}(t)&=&\frac{1}{2}m \dot\alpha^2 \pm f(t) x_0(t) \mp c \alpha(t). 
\eeqa
The above structure facilitates the formal solution of the dynamics as the time-dependent part only accumulates a phase, whereas the time-independent part gives a simple evolution operator, 
\beqa
\ket{\psi_I^{\uparrow \downarrow}(t)}&=&\exp (-\frac{i}{\hbar} \int_0^t{F^{\uparrow \downarrow}(t') dt'}) \ket{\psi_{I,0}^{\uparrow \downarrow}(t)}, \nonumber \\
\ket{\psi_{I,0}^{\uparrow \downarrow}(t)}&=&e^{-i H_{I,0}t/\hbar} \ket{\psi_{I,0}^{\uparrow \downarrow}(0)}. 
\eeqa
Due to $H_{I,0}$ and $\ket{\psi_{I,0}^{\uparrow \downarrow}(0)}=\ket{\Phi(0)}$ are spin-independent, $\ket{\psi_{I,0}^{\uparrow \downarrow}(t)}=\ket{\Phi(t)}=e^{-i H_{I,0}t/\hbar} \ket{\Phi(0)}$ is also a spin-independent vector. Since $e^{\mp i \alpha p/\hbar}$ shifts the position representation as $\bra{x} e^{\mp i \alpha p/\hbar} \ket{\Phi}=\Psi(x\mp\alpha)$, the branch wave functions in the laboratory frame are found using Eq. (\ref{IP}), 
\beq
\label{Sstates}
\psi^{\uparrow \downarrow}(x,t)=e^{\pm i m \dot \alpha x/\hbar} \exp(-\frac{i}{\hbar} \int_0^t{F^{\uparrow \downarrow}(t') dt'} ) \Phi(x\mp\alpha, t). 
\eeq
Specially, at final time $t_f$, 
\beq
\label{Sstatestf}
\psi^{\uparrow \downarrow} (x,t_f)= \exp(-\frac{im}{2\hbar} \int_0^{t_f}{\dot \alpha^2(t) dt}) 
\exp (\pm \frac{ic}{\hbar} \int_0^{t_f}{\alpha(t) dt}) \exp (\mp \frac{i}{\hbar} \int_0^{t_f}{x_0(t) f(t) dt}) \Phi(x,t_f).
\eeq
For $x_0$ constant the phase terms
\beq
\label{x0f_int}
\mp x_0 \int_0^{t_f}{ f(t) dt}=0,
\eeq
because of Eqs. (\ref{bc}) and (\ref{Newton}). Consequently, the complex overlap in (\ref{overlap}) takes the very simple form 
\beq
\bra{\psi^\downarrow(t_f)}\ket{\psi^\uparrow(t_f)}=\exp(\frac{2ic}{\hbar} \int_0^{t_f}{\alpha(t) dt}) .
\eeq
As we can see the phase differential,  $\Delta \varphi(t_f)=\frac{2c}{\hbar}\int_{0}^{t_{f}}\alpha(t)dt=S c$,  is proportional to $c$  with the controllable sensitivity $S=\frac{2}{\hbar}\int_{0}^{t_{f}}\alpha(t)dt$.
%
%

\begin{figure}[h]
\begin{center}
\scalebox{1}[1]{\includegraphics[scale=0.5]{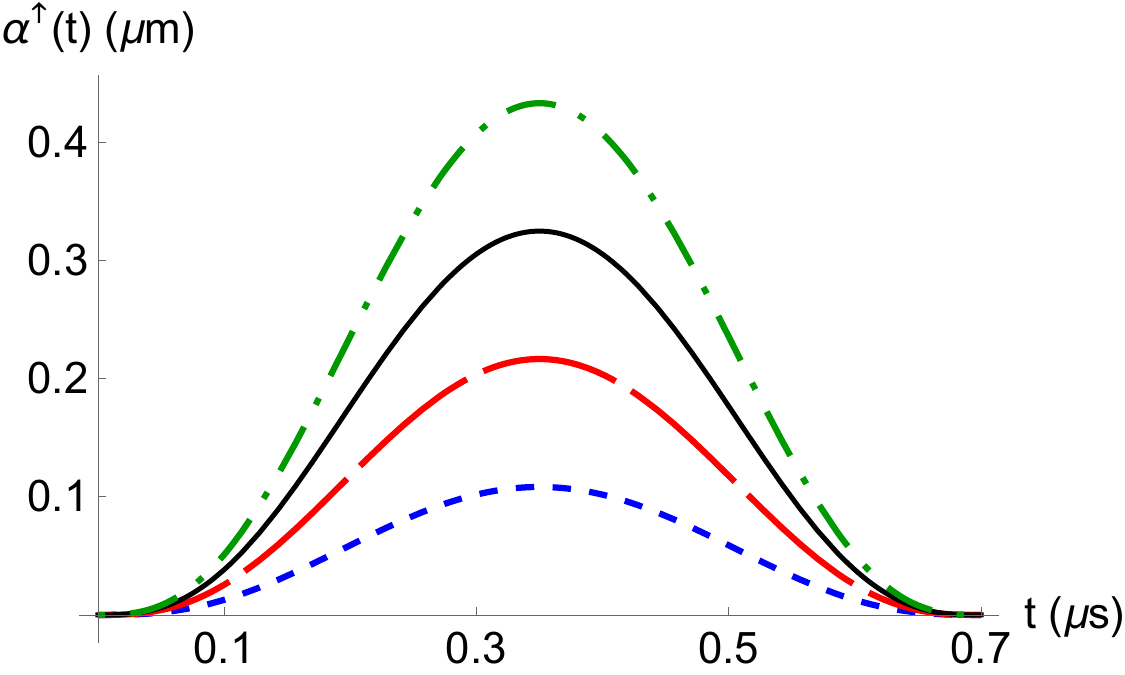}}
\caption{(Color
online) Trajectories at final time $t_{f}=0.7$ for different values of M. $M=\lambda /8$ (short dashed blue line), $M=\lambda$ (long dashed red line), $M=3 \lambda /8$ (solid black line) and $M=\lambda /2$ (dotdashed green line). Note: $\lambda=0.866$ $\mu$m.}
\label{Trajectories}
\end{center}
\end{figure}

{\bf Measurement of $c$ through populations.}
We first must set the trajectory $\alpha(t)$ , a particular solution of the Newton equation which satisfies the boundary conditions. We start with a sixth-order polinomial
\begin{equation}
\alpha(t)=\sum_{j=0}^{6}b_{j}\left(\frac{t}{t_{f}}\right)^j
\end{equation} 
and apply the required boundary conditions. We also impose an extra condition, $\alpha(\frac{t_{f}}{2})=M$, which implies  maximum displacement $M$  at $t=t_{f}/2$. 
\begin{equation}
\label{eq:trajectories}
\alpha(t)=64 M\left(\frac{t}{t_{f}}\right)^3-192 M\left(\frac{t}{t_{f}}\right)^4+192 M\left(\frac{t}{t_{f}}\right)^5-64 M\left(\frac{t}{t_{f}}.\right)^6
\end{equation}
So, by selecting a value of $M$ we set one trajectory, and we have one value of the sensitivity as $S=\frac{32 M}{35}\frac{t_{f}}{\hbar}$. From here, the driving force might be inverse engineered from the Newton equation (see \cite{Martinez-Garaot2018} and \cite{Rodriguez-Prieto2020}). 
In Figure \ref{Trajectories} we plot several trajectories for different values of the maximum displacement $M$.
%
%
%
%
\begin{figure}[h]
\begin{center}
\scalebox{1}[1]{\includegraphics[scale=0.5]{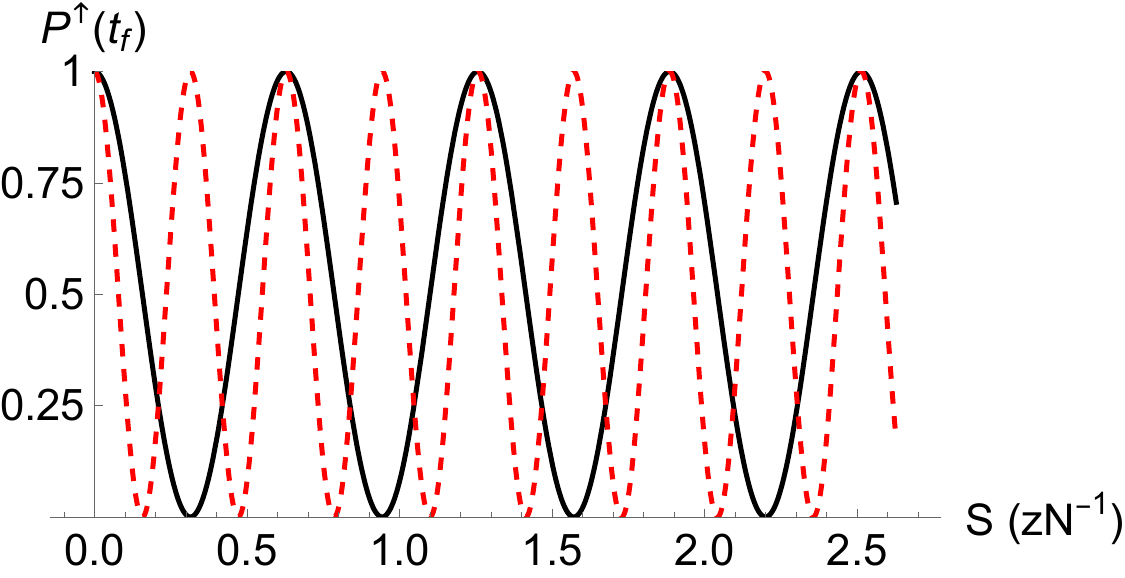}}
\caption{(Color
online) Spin-up state population after a $\pi/2$-pulse at final time $t_{f}=0.7$ $\mu$s as a function of the sensitivity of the interferometer $S=\frac{2}{\hbar}\int_0^{t_f} \alpha (t)dt$, which is varied changing the maximum displacement 
$M$ of the $\alpha(t)$ trajectories. $c=10$ zN (solid black line) and $c=20$ zN (dashed red line).}
\label{PopulationsNoNoise}
\end{center}
\end{figure}
%
%
Now, we can plot the populations $P^{\uparrow \downarrow}(t_f)=\frac{1}{2}\pm\frac{1}{2}\cos(c S)$ as a functions of $S$ (see Figure \ref{PopulationsNoNoise}), which are periodically oscillating functions. 
Therefore, we can extract $c$ from the oscillation period $ \pi\hbar/c$. Please note that the amplitude of the oscillation keeps constant as we have maximal visibility, this is, the modulus in Eq. (\ref{overlap}) is one. 

\section{Noise sensitivity}
Consider now that the force $f(t)$ may suffer from some noisy deviation from the ideal value. This deviation is represented by $\lambda \xi(t)$, possibly multiplied by some appropriate dimensional factor depending on the parameter. $\lambda$ is the dimensionless perturbative parameter that should be set to one at the end of the calculation, and $\xi(t)$ is also dimensionless. $\xi(t)$ is assumed to be unbiased, i.e., the average over noise realizations $\mathcal{E}[\dots]$ gives zero, and the (dimensionless) correlation function $\beta$ is stationary, 
\beq
\label{c_f}
\mathcal{E}[\xi(t)]=0, \quad \mathcal{E}[\xi(t)\xi(s)]=\beta(t-s). 
\eeq

For this kind of errors, the actual, experimentally implemented Hamiltonian is $H(t)=H_0(t)+\lambda H_1(t)$, being $H_0(t)$ the ideal unperturbed Hamiltonian. The evolution of the state is still described by the Schr\"odinger equation with the perturbed Hamiltonian $H(t)$, 
\beq
i \hbar \partial_t \psi(x,t)=[H_0(x,t)+\lambda H_1(x,t)] \psi(x,t). 
\eeq
Using perturbation theory up to $\mathcal{O}(\lambda^2)$, the state at final time $t_f$ can be written as \cite{Ruschhaupt2012}, 
\beqa
\label{pfinalstate}
\psi(x,t_f)=\psi_0(x,t_f)&-&i\frac{\lambda}{\hbar}  \int_0^{t_f} {dt U_0(t_f,t) H_1(x,t) \psi_0(x,t)} \nonumber \\
&-&\frac{\lambda^2}{\hbar^2}\int_0^{t_f}{dt \int_0^t{dt' U_0(t_f,t)H_1(x,t)U_0(t,t')H_1(x,t')\psi_0(x,t')}}+\dots
\eeqa
where $\psi_0(x,t)$ is the unperturbed solution and $U_0$ the unperturbed time evolution operator. 

In the case of the interferometer, each ``internal state''-dependent forces can suffer different noisy deviations, then the force perturbed by a realistic error can be described as
\beq
\label{noiseforce}
f(t)=f_\xi (t) [1+\lambda^{\uparrow \downarrow} \xi^{\uparrow \downarrow}(t)],
\eeq
%
so substituting (\ref{noiseforce}) in (\ref{Hinitial}) and assuming $x_0$ constant we get
\beq
\label{H1}
H_1^{\uparrow \downarrow}(x,t)=\mp (x-x_0) f_\xi(t) \xi^{\uparrow \downarrow} (t). 
\eeq

To see how the noise affect to the interferometer visibility we have to compute the overlap $\bra{\psi^\downarrow(t_f)}\ket{\psi^\uparrow(t_f)}$ in (\ref{overlap}). 
Due to each spin dependent forces can suffer different noisy deviations, according to Eq. (\ref{pfinalstate}), we define
\beqa
\psi^{\uparrow \downarrow}(x,t_f)=\psi_0^{\uparrow \downarrow}(x,t_f)&-&i\frac{\lambda^{\uparrow \downarrow}}{\hbar}  \int_0^{t_f} {dt U_0^{\uparrow \downarrow}(t_f,t) H_1^{\uparrow \downarrow}(x,t) \psi_0^{\uparrow \downarrow}(x,t)} \nonumber \\
&-&\frac{{\lambda^{\uparrow \downarrow}}^2}{\hbar^2}\int_0^{t_f}{dt \int_0^t{dt' U_0^{\uparrow \downarrow}(t_f,t)H_1^{\uparrow \downarrow}(x,t)U_0^{\uparrow \downarrow}(t,t')H_1^{\uparrow \downarrow}(x,t')\psi_0^{\uparrow \downarrow}(x,t')}} \nonumber \\
&+&\dots,
\eeqa
with $\psi_0^{\uparrow \downarrow}(x,t)$ the ones in (\ref{Sstates}). 
Therefore, taking into account the above equation, the overlap up to second order in $\lambda^{\uparrow\downarrow}$ is
\beqa
\label{overlap_3}
&&\bra{\psi^\downarrow(t_f)}\ket{\psi^\uparrow(t_f)}=\int_{-\infty}^\infty {dx {\psi_0^\downarrow}^\ast (x,t_f)\psi_0^\uparrow (x,t_f)} \nonumber \\
&-&i\frac{\lambda^\uparrow}{\hbar} \int_{-\infty}^\infty {dx \int_0^{t_f} {dt {\psi_0^\downarrow}^\ast (x,t_f) U_0^\uparrow(t_f,t) H_1^\uparrow(x,t) \psi_0^\uparrow (x,t)}} \nonumber \\
&-&\frac{{\lambda^\uparrow}^2}{\hbar^2} \int_{-\infty}^\infty{dx \int_0^{t_f} {dt \int_0^t {dt'  {\psi_0^\downarrow}^\ast (x,t_f) U_0^\uparrow (t_f,t) H_1^\uparrow(x,t)U_0^\uparrow(t,t') H_1^\uparrow(x,t') \psi_0^\uparrow(x,t')} } }  \nonumber \\
&+&i \frac{\lambda^\downarrow}{\hbar} \int_{-\infty}^\infty {dx \int_0^{t_f} {dt {\psi_0^\downarrow}^\ast (x,t){H_1^\downarrow}^\dagger (x,t) {U_0^\downarrow}^\dagger (t_f,t) \psi_0^\uparrow (x,t_f) } }  \nonumber \\
&+&\frac{\lambda^\downarrow \lambda^\uparrow}{\hbar^2} \int_{-\infty}^\infty {dx \int_0^{t_f} {dt {\psi_0^\downarrow}^\ast (x,t) {H_1^\downarrow}^\dagger (x,t) {U_0^\downarrow}^\dagger (t_f,t) U_0^\uparrow(t_f,t) H_1^\uparrow(x,t) \psi_0^\uparrow (x,t)} } \nonumber \\
&-&\frac{{\lambda^\downarrow}^2}{\hbar^2} \int_{-\infty}^\infty {dx \int_0^{t_f} {dt \int_0^t { dt' {\psi_0^\downarrow}^\ast (x,t') {H_1^\downarrow}^\dagger (x,t') {U_0^\downarrow}^\dagger (t,t') {H_1^\downarrow}^\dagger (x,t) {U_0^\downarrow}^\dagger (t_f,t) \psi_0^\uparrow (x,t_f)} } }. \nonumber \\
\eeqa
Paying attention to Eqs. (\ref{H1}) and (\ref{overlap_3}) it can be easily seen that some terms become zero when averaging over noise realizations due to $\mathcal{E}[\xi^{\uparrow \downarrow}(t)]=0$, then the above expression is reduced to
%
\beqa
&&\bra{\psi^\downarrow(t_f)}\ket{\psi^\uparrow(t_f)}=\int_{-\infty}^\infty {dx {\psi_0^\downarrow}^\ast (x,t_f)\psi_0^\uparrow (x,t_f)} \nonumber \\
&-&\frac{{\lambda^\uparrow}^2}{\hbar^2} \int_{-\infty}^\infty{dx \int_0^{t_f} {dt \int_0^t {dt'  {\psi_0^\downarrow}^\ast (x,t_f) U_0^\uparrow (t_f,t) H_1^\uparrow(x,t)U_0^\uparrow(t,t') H_1^\uparrow(x,t') \psi_0^\uparrow(x,t')} } }   \nonumber \\
&-&\frac{{\lambda^\downarrow}^2}{\hbar^2} \int_{-\infty}^\infty {dx \int_0^{t_f} {dt \int_0^t { dt' {\psi_0^\downarrow}^\ast (x,t') {H_1^\downarrow}^\dagger (x,t') {U_0^\downarrow}^\dagger (t,t') {H_1^\downarrow}^\dagger (x,t) {U_0^\downarrow}^\dagger (t_f,t) \psi_0^\uparrow (x,t_f)} } },\nonumber \\
\eeqa
or, using Eqs. (\ref{Sstates}), (\ref{Sstatestf}) and (\ref{H1}), 
\beqa
\label{overlap_2}
&&\bra{\psi^\downarrow(t_f)}\ket{\psi^\uparrow(t_f)}=\exp(\frac{2ic}{\hbar} \int_0^{t_f} {\alpha(t)dt}) \left \{ 1 \right. \nonumber \\
&-&\frac{{\lambda^\uparrow}^2}{\hbar^2} \int_0^{t_f} dt { \int_0^t dt' {f_\xi(t)f_\xi(t') \xi^\uparrow(t) \xi^\uparrow(t')  (x-x_0)^2 \int_{-\infty}^\infty {dx \Phi^\ast(x-\alpha,t) \Phi(x-\alpha,t)} } } \nonumber \\
&-&\left. \frac{{\lambda^\downarrow}^2}{\hbar^2} \int_0^{t_f} dt { \int_0^t dt' {f_\xi(t')f_\xi(t) \xi^\downarrow(t') \xi^\downarrow(t)  (x-x_0)^2 \int_{-\infty}^\infty {dx \Phi^\ast(x+\alpha,t) \Phi(x+\alpha,t)} } }  \right \}. \nonumber \\
\eeqa

We will consider now for simplicity the case in which the trap potentials are harmonic oscillators with frequency $\omega$ and states
\beq
\Phi_n(x\pm\alpha,t)=\frac{1}{\sqrt{2^nn!}} \left ( \frac{m \omega}{\pi \hbar} \right)^{1/4} e^{-\frac{m \omega}{2 \hbar}(x\pm \alpha)^2}H_n \left( \sqrt{\frac{m \omega}{\hbar}} (x\pm\alpha) \right ), 
\eeq
where $H_n$ are the Hermite polynomials. 
Assuming the initial state is the ground state, only the diagonal terms will be relevant. Therefore, Eq. (\ref{overlap_2}) becomes
\beqa
&&\bra{\psi^\downarrow(t_f)}\ket{\psi^\uparrow(t_f)}=\exp(\frac{2ic}{\hbar} \int_0^{t_f} {\alpha(t)dt}) \left \{ 1 \right. \nonumber \\
&-&\frac{{\lambda^\uparrow}^2}{\hbar^2} \int_0^{t_f} dt { \int_0^t dt' {f_\xi(t)f_\xi(t') \xi^\uparrow(t) \xi^\uparrow(t') \left [ \alpha^2(t)+(2(n+1)-1)\frac{\hbar}{2m\omega}-2x_0\alpha(t)+x_0^2 \right ]} } \nonumber \\
&-&\left. \frac{{\lambda^\downarrow}^2}{\hbar^2} \int_0^{t_f} dt { \int_0^t dt' {f_\xi(t')f_\xi(t) \xi^\downarrow(t') \xi^\downarrow(t) \left [ \alpha^2(t)+(2(n+1)-1)\frac{\hbar}{2m\omega}+2x_0\alpha(t)+x_0^2 \right ] } } \right \}. \nonumber \\
\eeqa

Averaging over different realizations of the noise and using (\ref{c_f})
\beqa
&&\mathcal{E}[\bra{\psi^\downarrow(t_f)}\ket{\psi^\uparrow(t_f)}]=\exp(\frac{2ic}{\hbar} \int_0^{t_f} {\alpha(t)dt}) \left \{ 1 \right. \nonumber \\
&-&\frac{{\lambda^\uparrow}^2}{\hbar^2} \int_0^{t_f} dt { \int_0^t dt' {f_\xi(t)f_\xi(t') \beta^{\uparrow}(t-t') \left [ \alpha^2(t)+(2(n+1)-1)\frac{\hbar}{2m\omega}-2x_0\alpha(t)+x_0^2 \right ]} } \nonumber \\
&-&\left. \frac{{\lambda^\downarrow}^2}{\hbar^2} \int_0^{t_f} dt { \int_0^t dt' {f_\xi(t')f_\xi(t) \beta^{\downarrow}(t'-t) \left [ \alpha^2(t)+(2(n+1)-1)\frac{\hbar}{2m\omega}+2x_0\alpha(t)+x_0^2 \right ] } } \right \}. \nonumber \\
\eeqa

To evaluate the integrals of the above expression we will assume white noise, $\beta^{\uparrow \downarrow}(t-t')=\gamma^{\uparrow \downarrow} \delta(t-t')$. Note that although the noises are independent for each branch they can have the same scale factor with the noise, so it seems logical to set $\gamma^\uparrow=\gamma^\downarrow=\gamma$ and $\lambda^\uparrow=\lambda^\downarrow=\lambda$. Therefore, the overlap becomes
\beqa
&&\bra{\psi^\downarrow(t_f)}\ket{\psi^\uparrow(t_f)}=\exp(\frac{2ic}{\hbar} \int_0^{t_f} {\alpha(t)dt}) \left \{ 1 \right. \nonumber \\
&-&\left. \frac{\lambda^2}{\hbar^2} \gamma \int_0^{t_f} {dt \left[ 2\alpha^2(t) + (2(n+1)-1)\frac{\hbar}{m\omega}+2x_0^2 \right] \int_0^t {dt' f_\xi(t')f_\xi(t) \delta(t'-t)} } \right \}. 
\eeqa
Under the convention $\int_0^t {dt' f(t') \delta(t'-t)}=\frac{1}{2}f(t)$, 
\beqa
\bra{\psi^\downarrow(t_f)}\ket{\psi^\uparrow(t_f)}&=&\exp(\frac{2ic}{\hbar} \int_0^{t_f} {\alpha(t)dt}) \left \{ 1 \right. \nonumber \\
&-&\left. \frac{\lambda^2}{\hbar^2} \gamma \int_0^{t_f} {dt \left[ 2\alpha^2(t) + (2(n+1)-1)\frac{\hbar}{m\omega}+2x_0^2 \right] f^2(t) } \right \}, 
\eeqa
or taking into account Eq. (\ref{Newton}), 
\beqa
\label{finaloverlap}
\bra{\psi^\downarrow(t_f)}\ket{\psi^\uparrow(t_f)}&=&\exp(\frac{2ic}{\hbar} \int_0^{t_f} {\alpha(t)dt}) \left \{ 1 \right. \nonumber \\
&-&\left. \frac{\lambda^2}{\hbar^2} \gamma m^2 \int_0^{t_f} {dt \left[ 2\alpha^2(t) + (2(n+1)-1)\frac{\hbar}{m\omega}+2x_0^2 \right] \ddot \alpha^2(t) } \right \}. 
\eeqa
%

%
%
\begin{figure}[h!]
\begin{center}
\scalebox{1}[1]{\includegraphics[scale=0.5]{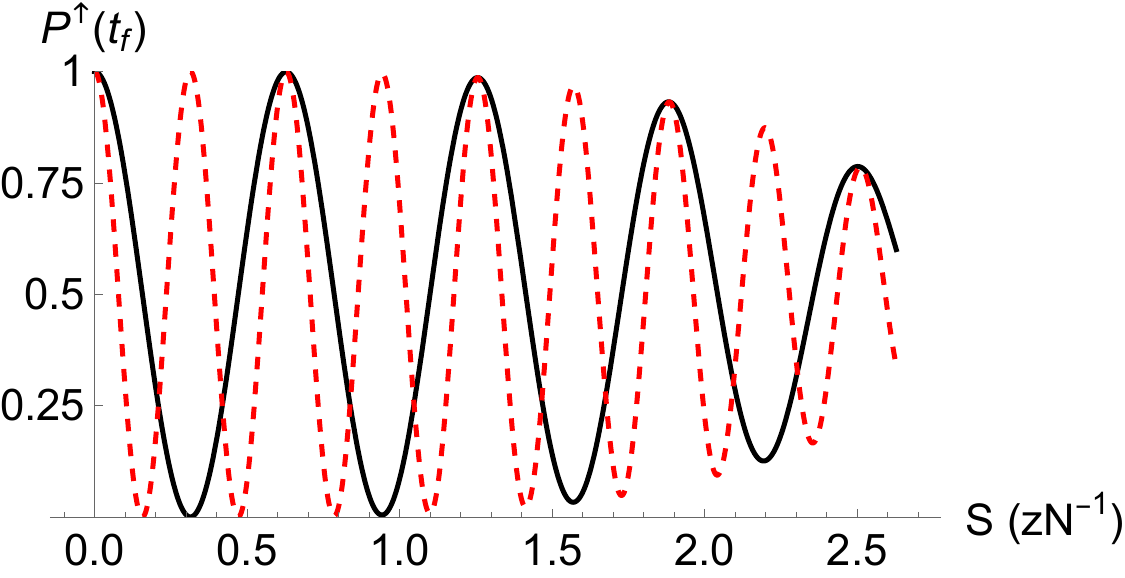}}
\caption{(Color
online) Spin-up state population after a $\pi/2$-pulse at final time $t_{f}=0.7$ $\mu$s as a function of the sensitivity of the interferometer $S=\frac{2}{\hbar}\int_0^{t_f} \alpha (t)dt$, which is varied changing the maximum displacement 
$M$ using $\alpha(t)$ trajectories. $c=10$ zN (solid black line) and $c=20$ zN (dashed red line).}
\label{PopulationsNoise}
\end{center}
\end{figure}
%

Since the terms that go with the error are real there will be no contribution of the error in the phase difference. Accordingly, the measure of the unknown force $c$ will not be affected. The visibility, instead, will be reduced as the modulus of the overlap is not 1. In fact, if we want to follow the same strategy commented in Section \ref{interferometer} to measure $c$, we need to plot populations as a function of the sensitivity (see Fig. \ref{PopulationsNoise}). As we can see in Figure \ref{PopulationsNoise}, the noisy terms affect  the amplitude of the oscillations which do not keep constant anymore. 

Analyzing in more detail the noisy terms in Equation (\ref{finaloverlap}), it is easy to see that, since $x_0$ is of order of c, the term proportional to $x_0^2$ is of second order and can be neglected in the weak-force regime. Moreover, $\sqrt{\hbar/m\omega}$ defines the characteristic length of the harmonic oscillator, i.e., the spatial width of the ground-state wave packet. As long as the displacement $\alpha (t)$ is significantly larger than this characteristic length, the dominant contribution to the visibility loss arises from the term $\alpha^2(t)\ddot{\alpha}^2(t)$.

\section{Optimal trajectories.}

In this section, we show how to design trajectories that minimize the effect of noise on visibility. As seen in Eq.~(\ref{finaloverlap}), the relevant quantity is
\begin{equation}
W = \int_0^{t_f} \alpha^2(t) \ddot{\alpha}^2(t) \, dt.
\end{equation}

Our goal is to maximize visibility (i.e., the modulus of Eq.~(\ref{finaloverlap}) as close to one as possible) while maintaining a fixed sensitivity
\begin{equation}
S = \frac{2}{\hbar} \int_0^{t_f} \alpha(t)\, dt = \Delta \varphi(t_f)/c.
\end{equation}

This leads to a constrained optimization problem: we aim to minimize the objective functional \( W \) under the integral constraint
\begin{equation}
S^* = \int_0^{t_f} \alpha(t)\, dt = \text{const}.
\end{equation}

The optimal trajectories \( \alpha(t) \) are solutions of the Euler–Poisson equation associated with the augmented Lagrangian
$L = \alpha^2 \ddot{\alpha}^2 - \Lambda \alpha$,
where $\Lambda$ is a Lagrange multiplier enforcing the isoperimetric constraint. The Euler–Poisson equation becomes:
\begin{equation} 
\label{eq:EP}
 \frac{\partial L}{\partial \alpha}-\frac{d}{dt}\frac{\partial L}{\partial \dot\alpha}+\frac{d^2}{dt^2}\frac{\partial L}{\partial \ddot\alpha}=0\Rightarrow 6\alpha \ddot\alpha^2
+4 \dot\alpha^2 \ddot\alpha
+8\alpha \dddot \alpha \dot\alpha
+2\alpha^2 \ddddot \alpha=\Lambda
\end{equation}

Imposing boundary conditions $\alpha(t_b) = \dot{\alpha}(t_b) = 0$ leads to the trivial solution $\alpha(t) = 0$, which has no physical interest. Moreover, the equation becomes singular at $\alpha(t) = 0$. To address this, we slightly shift the boundary conditions, setting $\alpha(t_b) = \epsilon$ and $\dot{\alpha}(t_b) = 0$, thereby avoiding the singularity. After solving the problem, we will analyze the limit \(\epsilon \to 0\) to assess the physical relevance of the results.

To facilitate numerical treatment, we introduce the dimensionless variables $\theta(\tau) = \frac{\alpha(t)}{\epsilon}$ and
$\tau = \frac{t}{t_f}$, resulting in the scaled differential equation:
\begin{equation}
\label{4rd_order_ODE_scaled}
6\theta \ddot{\theta}^2
+ 4 \dot{\theta}^2 \ddot{\theta}
+ 8\theta \dot{\theta} \dddot{\theta}
+ 2\theta^2 \ddddot{\theta}
= \frac{\Lambda t_f^4}{\epsilon^3},
\quad
\theta(\tau_b) = 1,\quad \dot{\theta}(\tau_b) = 0,
\end{equation}
where dots now denote derivatives with respect to \(\tau\), and the boundaries are fixed at \(\tau_0 = 0\), \(\tau_f = 1\).
This scaled problem depends only on a single dimensionless parameter $\delta = \frac{\Lambda t_f^4}{\epsilon^3}$.
The integration domain is always \(\tau \in [0, 1]\), and the results must be rescaled to physical units, specifically
$S^* = \epsilon t_f \tilde{S}$ and $\quad W = \frac{\epsilon^4}{t_f^3} \tilde{W}$, where the tilded quantities are dimensionless results from the scaled problem.

Eq.~(\ref{4rd_order_ODE_scaled}) is solved numerically using MATLAB’s \texttt{bvp4c} routine to obtain $\theta(\tau)$. From this, we compute
\begin{equation}
\tilde{S} = \int_0^1 \theta(\tau)\, d\tau, \quad \tilde{W} = \int_0^1 \theta^2(\tau) \ddot{\theta}^2(\tau)\, d\tau.
\end{equation}

\begin{figure}[t!]
\centering
\includegraphics[width=\textwidth]{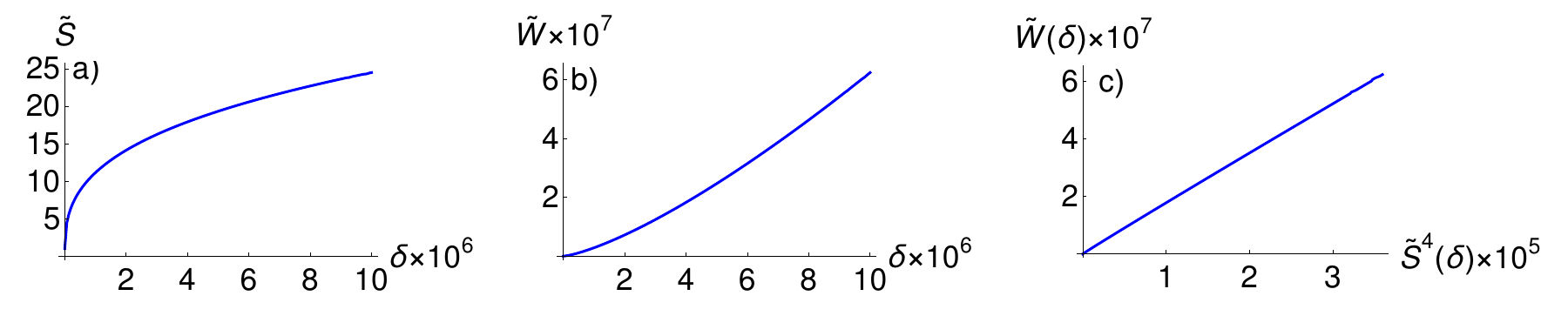}
\caption{(a) Dimensionless visibility $\tilde S=\int_0^1\theta(\tau) d\tau$ and (b) dimensionless effect of the noise $\tilde W=\int_0^1\theta(\tau)^2\ddot\theta(\tau)^2 d\tau$ as a function of $\delta$. 
In (c), $\tilde W(\delta)$ is plotted as a function of $\tilde S^4$ as $\delta$  varies parametrically between $(0,10^7)$, showing 
that the linear $\tilde W \propto\tilde S^4$ relation may be a good approximation. Using numerical polynomial fitting, a slope of 
$\frac{\tilde W}{\tilde S^4}\sim 174$ is obtained.}
\label{Optimization1}
\end{figure}

By varying $\delta$, we obtain a parametric plot of $\tilde{W}(\delta)$ versus $\tilde{S}(\delta)$, which characterizes the optimal trade-off between noise and sensitivity. Our numerical results suggest a quartic dependence $\tilde{W} \approx k \tilde{S}^4$,
with a numerically fitted constant $k \approx 174$, see Fig.~\ref{Optimization1}.

Returning to the original (non-scaled) formulation, and using the relations \( S = \epsilon t_f \tilde{S} \) and \( W = \frac{\epsilon^4}{t_f^3} \tilde{W} \), we can write
\begin{equation}
W = \left( \frac{\tilde{W}}{\tilde{S}^4} \right) \frac{S^4}{t_f^7} \approx 174 \, \frac{S^4}{t_f^7}.
\end{equation}

Remarkably, the dependence on $\epsilon$ cancels out on both sides of the expression. 
This result reveals a fundamental limitation: the dependence of $W$ on $S$ imposes a lower bound on the minimum achievable noise for a given sensitivity.

\subsection{Comparison of optimal and polynomial trajectories}
\label{comparison_sec}
 
\begin{figure}[h!]
    \centering
    \includegraphics[scale=0.75]{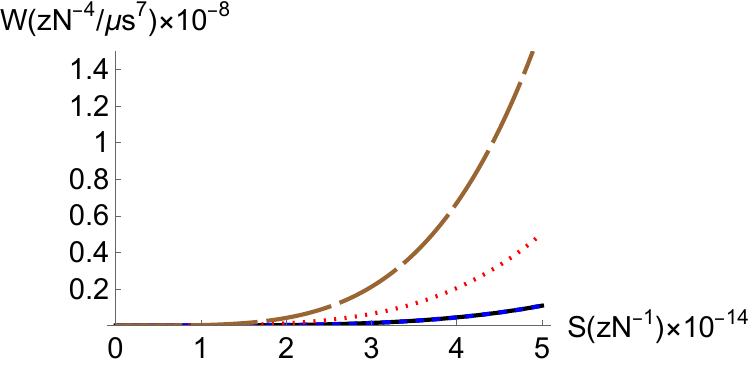}
    \caption{(Color online) Noise-sensitivity relation $W(S)$ for a final time $t_f=1\mu s$ in the four different scenarios analyzed in subsection \ref{comparison_sec}. Noise–sensitivity relation $W(S)$ for a final time $t_f = 1\,\mu\text{s}$, shown for the four trajectory scenarios analyzed in Subsection~\ref{comparison_sec}. From top to bottom: 
    sixth-degree polynomial trajectory $W = 2614 \frac{S^4}{t_f^7}$ (brown dashed),
    fourth-degree polynomial trajectory $W = 827 \frac{S^4}{t_f^7}$ (red dotted), 
    optimal trajectory $W = 174 \frac{S^4}{t_f^7}$ (black solid), and the 
    purely numerical bound $W_{\text{nb}} = \left(\frac{\tilde{W}}{\tilde{S}^4}\right)\frac{S^4}{t_f^7}$ (blue dashed). 
    Note that the optimal and numerical bounds are nearly indistinguishable at the scale of the figure.}
    \label{Optimization2}
\end{figure}

To conclude this section, we compare the optimal relation between $W$ and $S$ with the one obtained using a simple polynomial ansatz for the $\alpha(t)$ trajectory, as introduced in Eq.~(\ref{eq:trajectories}),
\begin{equation}
\alpha(t) = M\left(\frac{64 t^3}{t_f^3} - \frac{192 t^4}{t_f^4} + \frac{192 t^5}{t_f^5} - \frac{64 t^6}{t_f^6}\right),
\end{equation}
where $M$ is the maximum displacement. Computing $S$ and $W$ for this trajectory leads to
\begin{equation}
S = \int_0^{t_f} \alpha(t)\, dt = \frac{16}{35} M t_f, \quad 
W = \int_0^{t_f} \alpha^2(t)\, \ddot{\alpha}^2(t)\, dt = \frac{16777216}{146965} \frac{M^4}{t_f^3},
\end{equation}
so that the resulting relation is $W = 2614 \frac{S^4}{t_f^7}$. 
For completeness, we also analyze a trajectory based on a fourth-degree polynomial, which does not satisfy the boundary conditions on the second derivative. Specifically, we consider:
\begin{equation}
\alpha(t) = M\left(\frac{16 t^4}{t_f^4} - \frac{32 t^3}{t_f^3} + \frac{16 t^2}{t_f^2}\right),
\end{equation}
which leads to the following $W(S)$ relation $W = 827 \frac{S^4}{t_f^7}$.
The results obtained across these four different scenarios are compared in Fig.~\ref{Optimization2}.

%

\section{Discussion}
\label{discussion}
Our results confirm that STA-based atom interferometers can be made robust against certain classes of noise, particularly those affecting the compensating force f(t). As shown analytically and numerically, the phase difference between the interferometric branches—responsible for encoding the measured force—remains unaffected by weak, unbiased noise, in agreement with previous works \cite{Rodriguez-Prieto2020,Martinez-Garaot2018}. However, the modulus of the overlap is reduced, leading to a loss of visibility that limits the contrast of the interference fringes and, consequently, the precision of the measurement.

This visibility degradation is quantified through a perturbative expansion up to second order in the noise strength, revealing that the dominant contribution arises from the term $\alpha^2(t) \ddot \alpha^2(t)$, provided the displacement is large compared to the oscillator length $\sqrt{\hbar}/(m\omega)$. This observation is consistent with previous studies on noise-induced excitation in STA-driven transport \cite{Lu2014,Lu2018,Ruschhaupt2012}, and highlights the importance of trajectory design in minimizing decoherence effects.

To address this, we formulated and solved an optimization problem under an isoperimetric constraint, identifying the class of trajectories that minimize the noise-induced loss of visibility for a fixed sensitivity. The resulting optimal trajectories exhibit a quartic scaling law $W \propto S^4$, with a prefactor significantly smaller than that of standard polynomial ans\"atze. This scaling behavior provides a practical guideline for designing robust interferometric sequences.

Our approach complements recent developments in robust quantum control \cite{Saywell2023} and enhanced STA methods (eSTA) \cite{Hauck2022}, which aim to preserve fidelity under more general noise models and in complex potential landscapes. While our analysis focuses on white noise and harmonic traps, future work could extend this framework to colored noise, non-Gaussian fluctuations, or anharmonic potentials, as explored in \cite{Lu2018,Lu2020}. Moreover, experimental validation of the predicted scaling laws and optimal trajectories would be a valuable step toward implementing noise-resilient quantum sensors based on STA interferometry.

\section*{Acknowledgments}
This work was supported by the Basque Government through Grant No. IT1470-22 and by Grant No. PID2021-126273NB- I00 funded by MCIN/AEI/10.13039/501100011033.  
%
%

\bibliographystyle{quantum}
\bibliography{Bibliography}
\end{document}